\documentstyle[11pt]{article}
\begin{document}

\centerline{\large\bf A Comprehensive Four-Quark Interpretation of
$D_s(2317)$, $D_s(2457)$ and $D_s(2632)$}

\vspace{0.6cm}

\centerline{Yu-Qi Chen$^1$ and Xue-Qian Li$^2$}

\vspace{0.2cm}

1. Institute of Theoretical Physics, Chinese Academy of Sciences,
100080, Beijing, China.\\

2. Department of Physics, Nankai University, 300071, Tianjin,
China.

\vspace{0.5cm}

\begin{center}
\begin{minipage}{12cm}
\noindent Abstract

{\small The recently observed new member of the charm-strange
family $D_s(2632)$ which has a surprisingly narrow width is
challenging our theory.  $D_s(2317)$ and $D_s(2457)$ which were
observed earlier have similar behaviors and receive various
theoretical explanations. Some authors use the heavy hadron chiral
effective theory to evaluate heavy-light quark systems and obtain
a reasonable evaluation on the masses of $D_s(2317)$ and
$D_s(2457)$. An alternative picture is to interpret them as
four-quark or molecular states. In this work, we are following the
later and propose a unitive description for all the three new
members $D_s(2632)$, $D_s(2317)$ and $D_s(2457)$ and at least, so
far our picture is consistent with the data.}
\end{minipage}

\end{center}

\vspace{1cm}

1. Introduction

Recently, the SELEX collaboration reported a new observation of a
narrow resonance $D_s(2632)$ with a mass of $2632\pm 1.6\; {\rm
MeV/c^2}$ and total width $< 17 $ MeV \cite{SELEX}. This resonance
is the heaviest member, so far, of the charm-strange family.
$D_s(2632)$ and the other two resonances $D_s(2317)$ and
$D_s(2457)$, which were observed earlier\cite{BABAR,BELLE,CLEO},
constitute an exotic group which is obviously distinguished from
the ordinary $D_s(1968)$ and $D^*_s(2112)$ by the masses, narrow
total widths and decay modes. The total decay widthes of these
states seem to be too small to be understood in the regular $q\bar
q'$ structure. If the newly observed resonances $D_s(2632)$ were
of a simple quark structure of $c\bar s$, one would expect that
the heavier the resonance is, the shorter its lifetime should be,
but the observation seems not to follow this pattern. These
unusual features challenge our present theory.

There have been some theoretical models proposed to interpret the
structures of $D_s(2317)$ and $D_s(2457)$ which are most likely to
possess spin-parity as $0^+$ and $1^+$ respectively
\cite{Pompili,Porter}. For example, Cahn and Jackson
\cite{Cahn,DiPierro} studied the new forms of the spin-orbit and
tensor forces in the heavy-light quark system and applied the
technique to the $D_s(2317)$ meson. In the special models, a mass
difference was predicted which is close to the experimental data.
Especially, two groups independently  work on the spectra by using
the heavy hadron chiral effective theory \cite{Bardeen,Nowak}. In
their work, they study the heavy-light quark hadrons where the
doublet $[0^+,\;1^+]$ stands as a chiral partner of the doublet
$[0^-,\; 1^-]$. Thus the mass difference between the two doublets
can be accounted by the Godberg-Trieman relation and is roughly
$m_N/3$ where $m_N$ is the mass of nucleon. It is consistent with
the present experimental data. Instead, the authors of other
theoretical papers suggested that the newly observed $D_s(2317)$
and $D_s(2457)$ are not in the simple $c\bar s$ composition, but
possibly are four-quark states, or mixtures of four-quark states
and $c\bar s$ \cite{Barnes,Beveren,Cheng}.

The discovery of $D_s(2632)$ which is 650 MeV heavier than the
ground state $D_s$, indicates that all three $D_s(2317)$
$D_s(2457)$ and $D_s(2632)$ might constitute a new group which is
different from the normal members of the charm-strange family.
Thus we are motivated to search for a unitive picture to describe
all of the three new resonances. In this work we follow the
four-quark picture and see if it can fit the data. Concretely, it
is postulated that all the three resonances are four-quark states
and have special spin-parity structures with narrow total widths.
Since all of the resonances are isospin singlets, and possibly are
four-quark states, we would determine their SU(3) quark structures
according to the general principle. We stress that in this
picture, we postulate that all the quarks are in S-waves. Since
$c\bar s$ is an iso-singlet, thus to constitute a four quark state
which is still an iso-singlet, the extra $q\bar q$ constituent
must be  an iso-singlet. As well known, in the SU(3) framework, an
iso-singlet can have two independent components, i.e. $u\bar
u+d\bar d$ and $s\bar s$. Because the orbital angular momenta for
all constituents are zero, the dynamics (which is still not very
clear so far) permits one of the two components to dominate in a
physical resonance. We propose that in $D_s(2317)$ and
$D_s(2457)$, the  $u\bar u+d\bar d$ component dominates, namely
they are of structure $c\bar s(u\bar u+d\bar d)$, while for
$D_s(2632)$, the $s\bar s$ component dominates, so the composition
is mainly $c\bar ss\bar s$ which is exactly an isospin partner of
the former one. Of course, QCD dynamics allows reaction $s\bar
s\rightarrow u\bar u+d\bar d$, which can occur by exchanging only
one gluon if they are in color octet, and then might mix the two
structures. But generally as the  system is in a color-octet, the
interaction between the quark and anti-quark is repulsive and the
annihilation process is strongly suppressed by the wavefunction.
Thus the
mixing between the two states can be, in general, neglected.\\

2. The $D_s(2317)$ and $D_s(2457)$ states.

As proposed, $D_s(2317)$ and $D_s(2457)$ would have the quark
structure $c\bar s(u\bar u+d\bar d)$ which guarantees the
iso-singlet requirement. Generally, they can be in either $(c\bar
s)_8(u\bar u+d\bar d)_8$ or $(c\bar s)_1(u\bar u+d\bar d)_1$ where
the subscripts 1 and 8 refer to the color singlet and octet
respectively, and may also be a mixture of the two states. With a
color re-combination, the former one can be transformed into a
$(c\bar u(\bar d))_1\otimes(u(d)\bar s)_1$ structure, which is a
$DK$ molecule. The molecular structure was proposed to interpret
some mesons such as $f_0(980)$ and $a_0(980)$ \cite{Weinstein},
several states of charmonium spectroscopy \cite{Rujula} and
$\chi(3.86)$ \cite{Tuan}. Here we will show that $D_s(2317)$ and
$D_s(2457)$ can be  accommodated in the molecular structure.
Actually, $D_s(2317)$ may be a molecule of DK mesons, whereas
$D_s(2457)$ is a $D^*K$ molecule.

The data support the picture of a molecular structure. The masses
of $D_s(2317)$ and $D_s(2457)$ are very close and slightly below
the thresholds of
$$M_{D^0}+M_{K^+}\sim M_{D^+}+M_{K^0}\sim (2358\sim 2367)\; {\rm
MeV},\;\;\;\;{\rm and}\;\;\;\;M_{D^*}+M_{K}\sim (2501\sim
2508)\:{\rm MeV}.$$ This is consistent with the molecular picture
\cite{Weinstein}. When the mass is near the threshold, $c$ and
$\bar{u} (\bar{d})$ quarks tend to form a $D$ meson while $u (d)$
and $\bar{s}$ quarks tend to form a $K$ meson. The two clusters
may be separated by a sizable distance.

Moreover,  it is noticed that the mass differences follow the
relations
\begin{equation}
\label{mass} M_{D^*}-M_D\approx M_{D_s^*}-M_{D_s}\approx
M_{D_s(2457)}-M_{D_s(2317)}\sim 140\;{\rm MeV}.
\end{equation}
These relations can be understood in the molecular picture and
imply that $D_s(2317)$ and $D_s(2457)$ are indeed molecular states
of $DK$ and $D^*K$ in S-wave respectively. In a system where a
heavy flavor is involved, the mass difference arises from
spin-flip of the charm quark and is suppressed by $1/m_c$
according to the heavy quark effective theory (HQET) \cite{Isgur}.
In the molecular states $DK$ or $D^*K$ the K-meson is pseudoscalar
which is blind to the charm spin, the energy difference of
$D_s(2317)$ and $D_s(2457)$ should completely come from the mass
difference of $D$ and $D^*$. Thus the relations (\ref{mass}) is
nicely understood.

The main decay modes of $D_s(2317)$ is $D_s\pi$. Because of the
constraint of the final state phase space, $D_s(2317)\rightarrow
D+K$ is forbidden. The only available channel is
\begin{equation}
D_s(2317)\rightarrow D_s^++\pi^0,
\end{equation}
where the flavor is conserved. However, this decay mode does not
conserve isospin which must be retained in the processes governed
by strong interaction. A possible explanation is that the reaction
is realized through $D_s(2317)$ decay into $ D_s$ and a virtual
$\eta $. With a virtual $\eta -\pi$ mixing  it eventually comes to
$D_s+\pi$ \cite{Cheng}. Certainly, this process is highly
suppressed. This explains why the decay width is so narrow.

For $D_s(2457)$, its main channel is $D_s(2457)\rightarrow
D^*_s+\pi^0$, which is also an isospin violated process and
suffers from the same suppression as for $D_s(2317)\rightarrow
D_s^++\pi^0$.

The experiment \cite{CLEO} has observed the decay channel
$D_s(2457) \to D_s+\gamma$ with a small branching ratio. In our
picture, this process can happen via  $u\bar u + d \bar d$
annihilating into photons. One gluon or photon exchange between
two constituent mesons is necessary to make this process possible.
Therefore, the rate is small. There has also been a search for
$D_s+\pi\pi$ channel and a negative result was reported. In fact,
there are several suppression effects for the process. The decay
of $D_s(2457)$ which is believed as $1^+$, into $D_s\pi\pi$ or
$D^*_s\pi\pi$ is a P-wave reaction and moreover, it is a
three-body final state process, the P-wave suppression and the
final state phase space integration greatly reduce the decay rate.
\\

3.  $D_s(2632)$.

The newly observed resonance raises a challenge to the theory. In
our picture, it is the SU(3) partner of $D_s(2317)$ and its
peculiar narrow total width can be naturally understood.

$D_s(2632)$ is supposed to be  dominated by the $c\bar s(s\bar s)$
structure. Since $s\bar s$ can only reside in
$\eta,\;\eta',\;\phi$, one might suspect if  $D_s(2632)$ could be
a $D_s\eta$ molecule, however, it is unlikely because its mass is
100 MeV above the threshold of $D_s$ and $\eta$. There is no
suitable molecular structure which can fit this quark composition
(as long as the spin-parity of $D_s(2632)$ is $0^+$ (in analog to
$D_s(2317))$. Thus we suppose that $D_s(2632)$ is a pure
four-quark state.

An intuitive picture of this state could be that the $c$ quark
stays in the middle and the three $s$ quarks with gluon clouds
move around it. Its mass is $3m_s+m_c+\Delta E_B$ where $\Delta
E_B$ is the binding energy and can only be evaluated by invoking
concrete models. Since constituent mass of strange quark is about
150 MeV heavier than the constituent mass of u and d quarks, the
state of $c\bar s(s\bar s)$ should be roughly 300 MeV heavier than
$c\bar s(u\bar u+d\bar d)$, and this simple evaluation is
consistent with the measurement.

Because of the special structure, one can expect the total width
of $D_s(2632)$ to be small. The reason is the Pauli exclusive
principle. As well known, the destructive interference between two
anti-d-quarks in $D^+\rightarrow K\pi$ can remarkably reduce the
total width of $D^{\pm}$ and makes its lifetime to be 2.5 times
longer than that of $D^0$ \cite{Bigi,He}. The same mechanism may
apply here. The two anti-strange quarks in the initial state would
eventually join the other quarks ($c$ and $s$) to constitute final
mesons, and an interchange of the two anti-s-quarks  can result in
a minus sign or in other words, a destructive interference.

The most natural decay mode is $D_s(2632)\rightarrow D_s\eta$ with
the $s\bar{s}$ pair forming an $\eta$ and $c\bar s$ forming a
$D_s$ meson. With the $s\bar{s}$ annihilates into $u\bar {u} +
d\bar {d}$ pair, it can also decay to $D+ K$. This is a higher
order process, so that is suppressed. Moreover, the color matching
of the quarks (anti-quarks) in the mesons leads to an additional
suppression. The experimental data show that the $D K$ decay
channel is 6 times smaller than the $D_s
\eta$ channel and it is indeed understandable in our picture.\\

4. Predictions.

In this four-quark state picture, one straightforward prediction
is the existence of the vector partner of $D_s(2632)$.

By a simple $SU(3)_f$ manipulation, we can guess the mass of a new
$1^+$ resonance $D_{sj}^*$. The following relation
\begin{equation}
M_{D_s(2632)}-M_{D_s(2317)}=M_{D^*_{sj}}-M_{D_s(2457)},
\end{equation}
seems to hold and we can expect the mass of $D_{sj}^*$ to be
\begin{equation}
D_{sj}^*\approx 2770 \;{\rm MeV}/c^2.
\end{equation}
Since it is not a molecular state (similar to $D_s(2632)$), one
cannot expect to make a precise estimate on its mass, so that if a
measured mass is 50 MeV deviated from our predicted value, it is
not surprising.  $D_{sj}^*$ is a $1^+$ meson whose total width is
small, i.e. a narrow resonance and is of quantum number of $c\bar
s$ (or $\bar c s$). Its decay modes should be dominated by $D_s^*
\eta$ and $D^* K$. This narrow resonance should be tested in the
near future experiments.

Another interesting prediction is that these states should have
decay modes to $e^+e^-$ pair with the light quark pair inside the
hadron annihilates into two photons with relatively large
branching ratios. Specifically, we would consider $D_s(2632) \to
D_s^* e^+e^-$, $D_s(2317) \to D_s^* e^+e^-$ and $D_s(2457) \to D_s
e^+e^-$. In the normal $c\bar{s} $ meson, the branching ratios of
these decay modes are very tiny. The measurement on these decays
will be a crucial test to the four quark state explanation.\\

5. Conclusion.

In this letter, we propose a simple unitive picture for the three
new members of the $D_s$ family which are observed in recent
experiments. We consider $D_s(2317)$ and $D_s(2457)$ to be of
$c\bar s(u\bar u+d\bar d)$ quark composition and moreover
motivated by the data, it is supposed that the quarks (antiquarks)
are re-combined into molecular states. Namely, $D_s(2317)$ is a
molecular state of $DK$ while $D_s(2457)$ is a $D^*K$ molecular
state. By constrast, the newly observed $D_s(2632)$ is of the
$c\bar s(s\bar s)$ structure and it is exactly the partner of
$D_s(2317)$ of $0^+$.

In this picture, we have naturally explain the narrowness of the
widths of all the three resonances and discussed some decay modes.
We further predict another narrow resonance which is a $1^+$ meson
and its mass is located at about 2770 MeV with a narrow width, we
also predict that these states should have relatively large
branching ratios for decaying  into $D_s^{(*)}$ and a $e^+e^-$
pair.

The  four-quark pictures for the three states accommodate the
newly observed resonances $D_s(2317)$, $D_s(2457)$ and $D_s(2632)$
and everything seems to be consistent with all existing
experimental data. Once  those predictions are confirmed by
further experiments, the existence of the four quark structure can
be established. Then it may open a new page in the hadronic physics.

In this short note, we do not account for any concrete dynamics,
but only apply the $SU(3)$ flavor symmetry to analyze the quark
structure of all the states $D_s(2317)$, $D_s(2457)$ and
$D_s(2632)$ and try to understand their qualitative behaviors,
such as the very narrow total widths and some decay modes. Since
the dynamics is not involved, some details would be missing, as we
discussed in last section about our prediction. It is, of course,
interesting and compelling to pursue
this line by adding the dynamics in the calculations. \\

\noindent{\bf Acknowledgment:}

One of the authors (YQC) wishes to thank C. Liu for bringing this
subject to his attention. This work is partially supported by the
National Natural Science Foundation of China.

\end{document}